\documentclass[10pt,twocolumn]{article} 
\usepackage{amsmath,amsfonts}
\usepackage{algorithmic}
\usepackage{algorithm}
\usepackage{array}
\usepackage[caption=false,font=normalsize,labelfont=sf,textfont=sf]{subfig}
\usepackage{textcomp}
\usepackage{stfloats}
\usepackage{url}
\usepackage{verbatim}
\usepackage{graphicx}
\usepackage{cite}
\usepackage[colorlinks=true,linkcolor=black,anchorcolor=black,citecolor=black,filecolor=black,menucolor=black,runcolor=black,urlcolor=black]{hyperref}

\usepackage[margin=1in]{geometry}
\vspace{-30 cm}

\begin{document}


\title{Mathematical proof of errors in Capasso's excess noise factor formula for an $n$-step staircase multiplier}

\author{Ankitha E. Bangera \\
\\
\it{Department of Electrical Engineering,} \\
\it{Indian Institute of Technology Bombay,} \\
\it{Mumbai$-$400076, India} \\
E-mail: ankitha\_bangera@iitb.ac.in; ankitha.bangera@iitb.ac.in}
\date{}

\maketitle
\thispagestyle{empty}

\begin{abstract}
Solid-state devices such as multistep staircase avalanche photodiodes (APDs) are analogs to the photomultiplier tubes and are considered cascade amplifiers. The major source of internal noise in these APDs is due to the randomness in their stepwise impact ionization. Recent literature on staircase APDs by research groups such as Campbell and co-workers has reported the theoretical estimates of total excess noise factors using Capasso's excess noise factor formula. This formula is based on Friis' total noise factor formula for cascade networks. This article proves that Capasso's formula for a staircase APD erroneously considers the power gains in Friis' total noise factor formula as the gains. 
\end{abstract}

\section{Introduction}
\label{sec1}

Staircase avalanche photodiodes (APDs) are semiconductor (or solid-state) devices that are analogs to the photomultiplier tubes (PMTs) with deterministic amplification \cite{bib1,bib2,bib3,bib4,bib5,bib6,bib7}. Therefore, it can be considered as a cascade-amplifier. The irregularities in the stepwise impact ionization are a major source of internal noise in these APDs \cite{bib2,bib3,bib5,bib6,bib8,bib9,bib10}. Some noise models \cite{bib3,bib4,bib5,bib6,bib8,bib9,bib10} for these APDs are reported in the literature. However, Capasso and co-workers were the first to report the total excess noise factor formula for $n$-stage staircase multipliers in the early 1980's \cite{bib5,bib6}. To date, Capasso's excess noise factor formula is used to theoretically estimate the total excess noise factor values in staircase APDs \cite{bib2,bib3,bib8}. Moreover, this formula is often used to analyze and fit experimentally measured data \cite{bib2,bib3,bib8}. From the literature, it is also known that these noise models are based on Friis' noise factor (or noise figure) formulas for cascade networks \cite{bib8,bib9,bib11,bib12}. This article proves that Capasso's formula for a staircase APD with higher step counts is wrong and will estimate incorrect values.

\section{Results and Discussion} 
\label{sec2}

Cappaso's excess noise factor formula and its equivalents for an $n$-step staircase APD are given by equations~\eqref{eqn_1} \cite{bib3,bib5,bib6,bib8} and~\eqref{eqn_2} \cite{bib2,bib9,bib10}. However, both equations are equivalent because `$\delta$' implies $(1-p)$. 

\begin{equation}
\label{eqn_1}
\begin{aligned}
F_{\text{T}_n} &= 1+\sum_{i=1}^{n} \left[ \frac{1}{(2-\delta)^i}.\frac{\delta(1-\delta)}{(2-\delta)} \right] \\
&=1+\frac{\delta}{(2-\delta)} \left[1-(2-\delta)^{-n}\right] \\
\end{aligned}
\end{equation} 

where `$\delta$' is the fraction of electrons at the input of each step that do not impact-ionize. The mean staircase gain of these APD in terms of `$\delta$' is $\langle m_\text{S} \rangle =(2-\delta)^n$, where $(2-\delta)$ is the mean gain at the $x$-th step ($\langle m_x \rangle$). 

\begin{equation}
\label{eqn_2}
\begin{aligned}
F_{\text{T}_n} &=1+\frac{(1-p)}{(1+p)} \left[1-(1+p)^{-n} \right] 
\end{aligned}
\end{equation} 

where `$p$' is the fraction of electrons at the input of each step that impact-ionize (or the stepwise ionization probability). The mean staircase gain of these APDs in terms of `$p$' is $\langle m_\text{S} \rangle=(1+p)^n$, where $(1+p)$ is the mean gain at the $x$-th step ($\langle m_x \rangle$).

Other formulas equivalent to equations~\eqref{eqn_1} and~\eqref{eqn_2} are listed below as equations~\eqref{eqn_3} and~\eqref{eqn_4} \cite{bib3,bib4}. 

\begin{equation}
\label{eqn_3}
\begin{aligned}
F_{\text{T}_n} &=1+\frac{var(m_1)}{\langle m_1 \rangle^2} + \sum_{i=2}^{n} \left[ \frac{var(m_i)}{\langle m_x \rangle^2 \prod_{j=1}^{(i-1)} \langle m_j \rangle} \right] \\
&=1+\frac{p_1(1-p_1)}{(1+p_1)^2} + \sum_{i=2}^{n} \left[ \frac{p_x(1-p_i)}{(1+p_i)^2 \prod_{j=1}^{(i-1)} (1+p_j)} \right] \\
&=1+\frac{(1-\delta_1)\delta_1}{(2-\delta_1)^2} + \sum_{i=2}^{n} \left[ \frac{(1-\delta_i)\delta_i}{(2-\delta_i)^2 \prod_{j=1}^{(i-1)} (2-\delta_j)} \right] \\
\end{aligned}
\end{equation} 

where $var(m_x)$ is the variance of the stepwise gain of the staircase APDs; $\langle m_x \rangle$ is the mean of its stepwise gain; and $p_x=(1-\delta_x)$ is the stepwise ionization probability. The equation can be used when the stepwise gains and its stepwise ionization probabilities differ for each step. However, if the stepwise gains and its stepwise ionization probabilities are considered equal, then equation~\eqref{eqn_3} reduces to equations~\eqref{eqn_1},~\eqref{eqn_2}, or~\eqref{eqn_4}. 

\begin{equation}
\label{eqn_4}
\begin{aligned}
F_{\text{T}_n} &=1+\frac{var(m)}{\langle m \rangle (\langle m \rangle -1)} \left[1-\frac{1}{\langle m \rangle^n} \right] 
\end{aligned}
\end{equation}

Moreover, from the literature \cite{bib8,bib9} it is known that the equations~\eqref{eqn_1}$-$\eqref{eqn_4} are based on equation~\eqref{eqn_5}. 

\begin{equation}
\label{eqn_5}
F_{\text{T}_n} = F_1+\sum_{i=2}^{n}\left(\frac{F_i-1}{\prod_{j=1}^{(i-1)}M_j}\right)
\end{equation}

where `$x$' is the step count; $F_x$ is the stage-wise Friis' noise factors; $M_x$ is the stage-wise gains; and `$n$' is the total number of stages in the cascade network. 

However, the original formula for the total (excess) noise factor of an $n$-stage cascade network in terms of its stage-wise (stepwise excess) noise factors given by Friis \cite{bib11,bib12} and later standardized by IEEE \cite{bib13} is given by,

\begin{equation}
\label{eqn_6}
F_{\text{T}_n}^{\text{Friis}} = F_1^{\text{Friis}}+\sum_{i=2}^{n}\left(\frac{F_i^{\text{Friis}}-1}{\prod_{j=1}^{(i-1)}G_j}\right)
\end{equation}

where `$x$' is the step count; $F_x^{\text{Friis}}$ is the stage-wise Friis' noise factors; and $G_x$ is the stage-wise power gains; and `$n$' is the total number of stages in the cascade network.

The major error in equations~\eqref{eqn_1}$-$\eqref{eqn_4} is that the `power gains' in equation~\eqref{eqn_5} was erroneously considered as `gains.' This is proved in subsection~\ref{subsec2_1}.

\subsection{Mathematical proof that Capasso's formula and its equivalents are incorrect} 
\label{subsec2_1}

The stepwise excess noise factor expression for $n$-step staircase APDs in terms of their stepwise ionization probability $p_x$ can be written as,

\begin{equation}
\label{eqn_7}
\begin{aligned}
F_x &=1+\frac{(1-p_x)}{(1+p_x)} \left[1-(1+p_x)^{-1} \right] \\
&= 1+\frac{p_x(1-p_x)}{(1+p_x)^2}
\end{aligned}
\end{equation}

If all the steps impact-ionize with equal ionization probabilities `$p$,' then the stepwise excess noise factors for $n$-step staircase APDs will be, 

\begin{equation}
\label{eqn_8}
\begin{aligned}
F_x &=1+\frac{p(1-p)}{(1+p)^2}
\end{aligned}
\end{equation}

For a 1-step staircase APD, equation~\eqref{eqn_7} or \eqref{eqn_8} or its equivalents will be the formula for its total excess noise factor (which will also be its stepwise excess noise factor), similar to a single-stage cascade network where its total noise factor is equal to its stage-wise noise factor ($F_{\text{T}_1}^{\text{Friis}} = F_1^{\text{Friis}}$). However, this does not apply to staircase APDs with higher step counts ($n\geq2$).

From equation~\eqref{eqn_2}, the total excess noise factor for a 2-step staircase APD will be, 

\begin{equation}
\label{eqn_9}
\begin{aligned}
F_{\text{T}_2} &=1+\frac{(1-p)}{(1+p)} \left[1-(1+p)^{-2} \right] 
\end{aligned}
\end{equation} 

Substituting equation~\eqref{eqn_8} in equation~\eqref{eqn_5} while considering $n=2$, we get, 

\begin{equation}
\label{eqn_10}
\begin{aligned}
F_{\text{T}_2} &=1+\frac{p(1-p)}{(1+p)^2} + \frac{\left[ 1+\frac{p(1-p)}{(1+p)^2}+1 \right]}{(1+p)} \\
&= 1+\frac{p(1-p)}{(1+p)^2}+\frac{p(1-p)}{(1+p)^3} \\
&= 1+\frac{(1-p)}{(1+p)} \left[ \frac{p}{(1+p)} + \frac{p}{(1+p)^2} \right] \\
&=1+\frac{(1-p)}{(1+p)} \left[1-\frac{1}{(1+p)^2} \right] 
\end{aligned}
\end{equation}

We observe that equation~\eqref{eqn_10} is equal to equation~\eqref{eqn_9}, only when the `power gains' in equation~\eqref{eqn_6} are erroneously considered as the `gains.'

Similarly, from equation~\eqref{eqn_2}, the total excess noise factor for a 3-step staircase APD will be,

\begin{equation}
\label{eqn_11}
\begin{aligned}
F_{\text{T}_3} &=1+\frac{(1-p)}{(1+p)} \left[1-(1+p)^{-3} \right] 
\end{aligned}
\end{equation}

Substituting equation~\eqref{eqn_6} in equation~\eqref{eqn_4} while considering $n=3$, we get,

\begin{equation}
\label{eqn_12}
\begin{aligned}
F_{\text{T}_3} &=1+\frac{p(1-p)}{(1+p)^2} + \frac{\left[ 1+\frac{p(1-p)}{(1+p)^2}+1 \right]}{(1+p)} \\
&~~~+ \frac{\left[ 1+\frac{p(1-p)}{(1+p)^2}+1 \right]}{(1+p)^2}\\
&= 1+\frac{p(1-p)}{(1+p)^2}+\frac{p(1-p)}{(1+p)^3}+\frac{p(1-p)}{(1+p)^4} \\
&= 1+\frac{(1-p)}{(1+p)} \left[ \frac{p}{(1+p)} + \frac{p}{(1+p)^2} + \frac{p}{(1+p)^3} \right] \\
&=1+\frac{(1-p)}{(1+p)} \left[1-\frac{1}{(1+p)^3} \right] 
\end{aligned}
\end{equation} 

We again observe that equation~\eqref{eqn_12} is equal to equation~\eqref{eqn_11}, only when the `power gains' in equation~\eqref{eqn_6} are erroneously considered as the `gains.' This error applies to all the higher step counts. 

Considering an $n$-step staircase APD, its total excess noise factor will be, 

\begin{equation}
\label{eqn_13}
\begin{aligned}
F_{\text{T}_n} &=1+\frac{p(1-p)}{(1+p)^2} + \frac{\left[ 1+\frac{p(1-p)}{(1+p)^2}-1 \right]}{(1+p)} \\
&~~~+ \frac{\left[ 1+\frac{p(1-p)}{(1+p)^2}-1 \right]}{(1+p)^2} \\
&~~~+...+\frac{\left[ 1+\frac{p(1-p)}{(1+p)^2}-1 \right]}{(1+p)^n} \\
&= 1+\frac{p(1-p)}{(1+p)^2}+\frac{p(1-p)}{(1+p)^3}+\frac{p(1-p)}{(1+p)^4} \\
&~~~+...+\frac{p(1-p)}{(1+p)^{n+1}} \\
&= 1+\frac{(1-p)}{(1+p)} \Biggl[ \frac{p}{(1+p)} + \frac{p}{(1+p)^2} + \frac{p}{(1+p)^3} \\
&~~~+...+ \frac{p}{(1+p)^n} \Biggr] \\
&=1+\frac{(1-p)}{(1+p)} \left[1-\frac{1}{(1+p)^n} \right] 
\end{aligned}
\end{equation} 

Again, for an $n$-step staircase APD, equation~\eqref{eqn_13} is equal to equation~\eqref{eqn_2} and its equivalent forms (equations~\eqref{eqn_2}$-$~\eqref{eqn_4}). But, since equation~\eqref{eqn_13} is derived from equation~\eqref{eqn_5}, which erroneously considers `power gains' in equation~\eqref{eqn_6} as `gains,' we conclude that equations~\eqref{eqn_1} to~\eqref{eqn_4} are incorrect.

We further demonstrate that the total excess noise factor values with $n\geq2$ for staircase APDs obtained using equation~\eqref{eqn_2} are not equal to those estimated using Friis' formula for cascade networks (equation~\eqref{eqn_6}). This is illustrated in Illustrations 1 to 3 in Appendix~\ref{secA1}. Further, Illustrations 2b and 3b in Appendix~\ref{secA1} prove that the equation~\eqref{eqn_2} is based on Friis' formula, where the power gains in Friis' equation~\eqref{eqn_6} were mistaken as the gains.

\section{Conclusion} 
\label{sec3}
 
We mathematically proved that Capasso's excess noise factor formula and its equivalent forms for an $n$-step staircase APD given by equations~\eqref{eqn_1}$-$~\eqref{eqn_4} erroneously consider the `power gains' in the original formula for a cascade network's total noise factor given by Friis as the `gains.' The evidence provided in the `Results and discussion' section suggests that Capasso's excess noise factor formula and its other forms for $n$-step staircase APDs reported in the literature \cite{bib2,bib3,bib4,bib5,bib6,bib8,bib9,bib10} are wrong. Thus, all the theoretical estimates of a staircase APD's total excess noise factors for higher step counts provided in the literature \cite{bib2,bib3,bib8} are also incorrect. This necessitates a correction to the existing formulas for staircase APDs. Our future work would be to derive a noise model for staircase APDs using a statistical method of a random process and extract the correct formula for the staircase APD's excess noise factor.

\section*{Acknowledgments}
A.E.B. would like to thank the Indian Institute of Technology Bombay for their support.

\section*{Appendices}
\appendix

\section{Illustrations: Mismatch in the estimated noise factor values} 
\label{secA1}

Friis' total noise factor formula for cascade networks \cite{bib11,bib12} is given by,

\begin{equation}
\label{Appendix_A_eqn_1}
F_{\text{T}_n}^{\text{Friis}} = F_1^{\text{Friis}}+\sum_{i=2}^{n}\left(\frac{F_i^{\text{Friis}}-1}{\prod_{j=1}^{(i-1)}G_j}\right)
\end{equation}

where $F_x^{\text{Friis}}$ is the Friis' noise factor at the $x$-th stage of cascade network (or stage-wise Friis' noise factors); $G_x$ is the power gain at the $x$-th stage of cascade network (or stage-wise power gains). For a single-stage network, the equation~\eqref{Appendix_A_eqn_1} reduces to, 

\begin{equation}
\label{Appendix_A_eqn_2}
F_{\text{T}_1}^{\text{Friis}} = F_1^{\text{Friis}}
\end{equation}

According to the literature \cite{bib8,bib9}, the previous noise models' total excess noise factor expressions are derived from the Friis' total noise factor formula for cascade networks (equation~\eqref{Appendix_A_eqn_1}). If `$p$' is the ionization probability at each step, then the total excess noise factor for $n$-step staircase APDs is given by, 

\begin{equation}
\label{Appendix_A_eqn_3}
\begin{aligned}
F_{\text{T}_n} &=1+\frac{(1-p)}{(1+p)} \left[1-(1+p)^{-n} \right] 
\end{aligned}
\end{equation} 

Here, the mean staircase gain in terms of `$p$' will be $\langle m_\text{S} \rangle =(1+p)^n$. The mean stepwise gains will be $M_x=\langle m_x \rangle = (1+p)$. \\

Bellow are some illustrations that estimate the values of stepwise and total excess noise factors for staircase APDs using equation~\eqref{Appendix_A_eqn_3} and compares these values with those estimated using Friis' noise factor formulas given by equations~\eqref{Appendix_A_eqn_1} and \eqref{Appendix_A_eqn_2}. Considering equal stepwise ionization probabilities as `$p$' at all steps of an $n$-step staircase APD, its stepwise excess noise factors will also be equal at all the steps (\textit{i.e.}, $\forall x$; $F_x=F$). \\

Let us consider the presence of internal noise, \textit{i.e.}, when the stepwise ionization probabilities `$p$' are less than unity and non-zero. \\

\textbf{Illustration 1:} If $p=0.3$ and $n=1$, then substituting the values of `$p$' and `$n$,' we get the total excess noise factor for 1-step staircase APD as $F_{\text{T}_1}=1.12426$. However, since this corresponds to a staircase APD with only one step, we consider that the stepwise excess noise factor at step one of this 1-step staircase APD as $F_1=F=F_{\text{T}_1}=1.12426=F_{\text{T}_1}^{\text{Friis}}$. \\

\textbf{Illustration 2a:} Similarly, if $p=0.3$ and $n=2$, the total excess noise factor for 2-step staircase APD will be $F_{\text{T}_2}=1.219845$. Since we have considered the same values of `$p$' as in the previous illustration, we have the stepwise excess noise factors at steps 1 and 2 as $F_1=F_2=F=1.12426$. Substituting these values in Friis' equation~\eqref{Appendix_A_eqn_1}, we get $F_{\text{T}_2}^{\text{Friis}} = F_1 + \frac{(F_2-1)}{G_1} = F_1 + \frac{(F_2-1)}{(1+p)^2} = 1.197787 \neq F_{\text{T}_2}$. \\

\textbf{Illustration 2b: Erroneously, stage-wise gain $M_x$ is considered in place of stage-wise power gain $G_x=M_x^2$:} Now, we get $F_{\text{T}_2}^{\text{Friis}} = F_1 + \frac{(F_2-1)}{M_1} = F_1 + \frac{(F_2-1)}{(1+p)} = 1.219845 = F_{\text{T}_2}$. \\

\textbf{Illustration 3a:} Similarly, if $p=0.3$ and $n=3$, the total excess noise factor for 3-step staircase APD will be $F_{\text{T}_3}=1.293372$.  Again since we have considered the same values of `$p$' as in the previous illustration, we have the stepwise excess noise factors at steps 1,2, and 3 as $F_1=F_2=F_3=F=1.12426$. Substituting these values in Friis' equation~\eqref{Appendix_A_eqn_1}, we get $F_{\text{T}_3}^{\text{Friis}} = F_1 + \frac{(F_2-1)}{G_1} + \frac{(F_3-1)}{G_1G_2} = F_1 + \frac{(F_2-1)}{(1+p)^2} + \frac{(F_3-1)}{(1+p)^2(1+p)^2} = 1.241294 \neq F_{\text{T}_3}$. \\

\textbf{Illustration 3b: Erroneously, stage-wise gain $M_x$ is considered in place of stage-wise power gain $G_x=M_x^2$:} Now, we get $F_{\text{T}_3}^{\text{Friis}} = F_1 + \frac{(F_2-1)}{M_1} + \frac{(F_3-1)}{M_1M_2} = F_1 + \frac{(F_2-1)}{(1+p)} + \frac{(F_3-1)}{(1+p)(1+p)} = 1.293372 = F_{\text{T}_3}$. \\

\end{document}